\documentstyle[12pt]{article}
\def\np{\vfill\eject}       
\def\ns{\vskip 2pc\noindent}       

\def\ni{\noindent}
\def\IR{I\kern-.255em R}
\def\cd{\cdot}
\def\part{\partial}
\def\siml{\underline \sim\,}

\def\la{\lambda}
\def\eps{\epsilon }
\def\epshl{{\epsilon\over 2} }

\def\yv{\vec{y} }                                                      
                                                      
\def\ev{\vec{e}}
\def\rv{\vec{r}}
\def\uv{\vec{u}}
\def\xv{\vec{x}}

\def\xvh{\hat{\vec{x}}}

\def\Db{{\bf D}\,}

\def\Fb{{\bf F}\,}
\def\Ga{{\bf \Gamma}\,}
\def\Gb{{\bf G}\,}
\def\Hb{{\bf H}\,}
\def\Ib{{\bf I}\,}
\def\Jb{{\bf J}\,}
\def\Kb{{\bf K}\,}

\def\Mb{{\bf M}\,}
\def\Pb{{\bf P}\,}
\def\Pd{\dot{\bf P}\,}

\def\Pbt{\tilde{\bf P}\,}
\def\Qb{{\bf Q}\,}
\def\Rb{{\bf R}\,}
\def\Sb{{\bf S}\, }

\def\La{{\bf \Lambda}}

\begin{document}
\begin{center}
{\bf BLOCK DIAGONALLY DOMINANT POSITIVE DEFINITE \\
APPROXIMATE FILTERS AND SMOOTHERS}	
\end{center}

{\it Running title:}

{\bf BLOCK DIAGONALLY DOMINANT APPROXIMATE FILTERS}	

Subtitle: Kalman filters and smoothers are approximated when the
transition matrix and the incremental information are nearly
block diagonal.
\medskip
\begin{center}
{\bf Kurt S. Riedel \\
Courant Institute of Mathematical Sciences \\
New York University \\
251 Mercer St. \\
New York, New York 10012}
\end{center}
\medskip

\begin{abstract}
We examine stochastic dynamical systems where the transition matrix,
$\Phi$, and the system noise, $\Ga\Qb\Ga^T$, covariance are nearly block
diagonal. When $\Hb^T \Rb^{-1} \Hb$ is also nearly block diagonal, where
$\Rb$ is the observation noise covariance and $\Hb$ is the observation
matrix, our suboptimal filter/smoothers are always positive semidefinite,
and have improved numerical properties. 
Applications for distributed dynamical systems 
with time dependent pixel imaging are discussed.
\end{abstract}
\newpage
{\bf I. INTRODUCTION}

In this article, we examine suboptimal filters and smoothers of
stochastic systems when the dynamics and the measurements are nearly
block diagonal (N.B.D.). We assume that the transition matrix, $\Phi (i+1,i)$,
the system noise covariance, $[\Ga\Qb\Ga^T]_i$, the initial state covariance,
$\Pb(0|0)$, and the measurement information matrix, $\Jb_i \equiv \Hb_i^T
\Rb_i^{-1} \Hb_i$, are all N.B.D. 
We then derive
estimation equations for the state vector, $\hat{\vec{x}}_i$,
and the covariance, $\Pb(i|i)$, which approximate the optimal
estimates to second order in $\eps$.

Our stochastic systems are similar to the widely studied weakly
coupled system (Kokotovic et al. (1969), 
Sezer and Siljak (1986), Gajic et al. (1990), Shen and Gajic (1990)).
Our N.B.D. systems are not limited to two block systems, but apply to an
arbitrary number of blocks. Furthermore, we require only
that $\Hb_i^T \Rb_i^{-1} \Hb_i$ is N.B.D. This contrasts to the
stronger hypothesis of weakly coupled systems that $\Hb_i$ and $\Rb_i$
are separately  weakly coupled.

The existing theory of weakly coupled systems concentrates on the
convergence of approximations to the complete system as $\epsilon$
tends to zero. Thus the existing analysis considers only the case
where $\epsilon$ is sufficiently small as to preclude the loss of
positive definiteness in the approximate equations.
Therefore previous analyses have not explicitly required positive
definiteness.

Our emphasis is on well-conditioned approximation of $\hat{\vec{x}}_i$
and $\Pb(i|i)$ for finite, but small values of the coupling parameter,
$\epsilon$.
Formally, our expansions require that the zeroth order  N.B.D. matrices
are all uniformly much larger than the remaining offdiagonal terms.
In practice, the coupling parameter, $\epsilon$, is not vanishingly small, and
there may be component directions where the first 
order terms almost cancel the zeroth order terms.
To prevent the approximate covariance matrix, $\Pb^{(\eps)}(i|i)$,
from losing positive definiteness, we add second order terms to the
approximate covariance. These additional terms not only guarantee
positive semidefiniteness, but also provide a matrix factorization.
 

Our motivation for the study of N.B.D. systems is the
analysis of distributed systems of partial differential equations
for fluid flow. We estimate the fluid flow as a function of time
and space, $\uv(\rv,t)$, where 
$\uv(\rv,t)$ satisfies the Navier-Stokes 
equations: $\part_t \uv + \uv\cdot \nabla \uv =\nabla p + \nu\Delta \uv$, 
$\nabla \cdot \uv= 0.$ 
We are given continuous time measurements of velocity field 
on a coarse grid in space. We expand the Navier-Stokes equation in
the set of eigenfunctions of the laminar flow linear stability problem
(Canuto et al. (1988)).
We truncate the eigenfunction expansion in the middle of the inertial
range, and model the effects of the discarded modes through an anomalously
large diffusion coefficient. 
 
When spatial inhomogeneities and nonlinearities are weak, 
the transition matrix, $\Phi (i+1,i)$, for the zeroth order eigenfunction
basis often will be N.B.D. 
To decouple the estimation equations to
leading order, we assume both $[\Ga\Qb\Ga^T]_i$ and $\Pb(0|0)$ are nearly block
diagonal.

In our prototypical system, the
elements of the measurement evaluation matrix, $\Hb_i$, are evaluations of
basis functions, $\Psi_k$, at the spatial locations, $z_{\ell}$, of
the measurements, $y_{\ell}$. 
When the $z_{\ell}$ are distributed more or less uniformly in space, and the
$\Psi_k$ are orthogonal, and $\Rb_i$ is a multiple, $\sigma_i^2$, of the
identity, then
$$(\Hb_i^T \Rb_i^{-1} \Hb_i )_{k,k'} =
{1\over \sigma_i^2}\sum_{\ell =1}^m \Psi_k (z_{\ell} ) \Psi_{k'} (z_{\ell}) 
\siml {m\over \sigma_i^2}
\int \Psi_k (z) \Psi_{k'} (z)dz \siml c_k \delta_{k,k'} .\eqno (1.1)
$$
 Thus for distributed systems of partial
differential equations with leading order to eigenfunctions as the basis
functions, the requirement that $\Hb_i^T \Rb_i^{-1} \Hb_i$ is nearly block
diagonal corresponds to the measurement locations being nearly uniformly
distributed and approximating spatial integration on the scalelength of the
shortest wavelength basis function. 

For such pixel type measurements,
the number of pixels needs to exceed the number of different diagonal blocks 
of eigenfunctions. If the measurements are spatially uniform, but
of insufficient number to distinguish the various eigenfunctions,
the evolution equations will be partially coupled due to spatial
aliasing.

Section II defines N.B.D. matrices, and presents 
several stabilizing transformations and approximate factorizations.
In Section III, we review the standard discrete Kalman filter and
derive a positive definite suboptimal approximation to the Kalman
filter. 
Section IV and Appendix B  derive similar suboptimal 
positive definite approximations to
the discrete Kalman smoothers for fixed intervals and for  fixed lags 
respectively.
Section V discusses our N.B.D. formulation.
Appendix A examines the
numerical advantages of computing the basic matrix operations
only to first order.

\medskip
\noindent
{\bf II.  NEARLY BLOCK DIAGONAL MATRIX
REPRESENTATIONS AND OPERATIONS}

A) {\it Matrix Structure}

We consider the class of nearly block diagonal (N.B.D.) matrices  to be
$N \times N$ matrices of the form:
$\Pb{(\eps)} = \Pb^{(0)}(\eps) + \eps \Pb^{(1)} +
\epsilon^2 \Pb^{(2)} (\epsilon) ,$ 
where $\Pb^{(2)}(\eps)$ contains second order and higher terms in $\eps$.
The weak coupling parameter, $\epsilon$, is a formal
small expansion term parameter. 
$\Pb^{(0)}(\eps)$ is block diagonal of the form:
$$
\Pb^{(0)}(\eps) = \left( \begin{array}{ccc}
\Pb_{11}^{(0)}(\eps)  & 0 & \ldots \\
0 & \Pb_{22}^{(0)}(\eps)  & 0 \\
0 & \ldots & 0 \\
0 & \ldots & \Pb^{(0)}_{N_b N_b}(\eps)  \end{array} \right) \ , \eqno (2.1)
$$
where the $\Pb_{kk}^{(0)}$ entry is a $n_k \times n_k$ matrix for
$k = 1,2, \ldots , N_b$. 
The block sizes,  $n_1 ,
n_2 , \ldots , n_{N_b}$, are fixed in this article, i.e.
all matrices have the same block structure.
We often suppress the functional dependence on $\eps$ in 
$\Pb^{(i)}(\eps)\ , i=0,1,2$.
We denote the truncated approximations of $\Pb(\eps)$ by $\Pb^{(\eps)}$,
where $\Pb^{(\eps)} =\Pb^{(0)} +\eps\Pb^{(1)}$ for first order approximations
and $\Pb^{(\eps)} =\Pb^{(0)} +\eps\Pb^{(1)}+\eps^2\Pb^{(2)}$ 
for second order approximations.

The first order block diagonal terms may be included in either
$\Pb^{(0)}(\eps)$ or $\eps\Pb^{(1)}$.
Including the block diagonal terms in $\Pb^{(0)}(\eps)$ reduces 
storage requirements; however, the resulting equations are slightly
more complicated. For simplicity, we include the first order block diagonal
terms in $\eps\Pb^{(1)}$.
We define $\Pb_L$ to be the strictly lower triangular part of a matrix $\Pb$
plus half of the block diagonal part of $\Pb$.


B) {\it Stabilizing Transformations and the $LD^{-1}L^T$ Factorization}

The truncated approximations, $\Pb^{(\eps)}$, to $\Pb(\eps)$ need not be
positive semidefinite even when $\Pb^{(0)}$ and $\Pb(\eps)$ are positive
definite. We assume that $\Pb^{(0)}$ is positive definite and 
that $\Pb^{(0)}$, $\Pb^{(\eps)}$ and $\Pb^{(\eps)}$ are symmetric,
we define the following transformations: 
$$ T_1[\Pb^{(\eps)}] \equiv 
[\Pb^{(0)} +\eps \Pb_L^{(1)}] \Pb^{(0) -1} [\Pb^{(0)} + \eps\Pb_L^{(1)} ]^T 
, \eqno (2.2) $$
for first order approximations, and
$$ T_2[\Pb^{(\eps)}] \equiv 
\left[\Pb^{(0)} +\eps \Pb_L^{(1)}+\eps^2 \left(\Pb_L^{(2)}- 
\Gb^{(2)}_L\right)\right] 
\Pb^{(0) -1} 
\left[\Pb^{(0)} +\eps \Pb_L^{(1)}+\eps^2 \left(\Pb_L^{(2)}- 
\Gb_L^{(2)}\right)\right]^T 
, \eqno (2.3) $$
for second order approximations where 
$\Gb^{(2)}\equiv \Pb_L^{(1)}\Pb^{(0) -1}\Pb_L^{(1)T}$.
Both transformations produce positive semidefinite matrices with
$LD^{-1}L^T$ block factorizations. $T_1[\Pb^{(\eps)}]$ differs from 
$\Pb^{(0)} +\eps \Pb^{(1)}$ by the second order term
$\eps^2 \Pb_L^{(1)}\Pb^{(0) -1}\Pb_L^{(1) T}$, and is thus strictly larger 
than $\Pb^{(0)} +\eps \Pb_L^{(1)}$. $T_2[\Pb^{(\eps)}]$ differs from 
$\Pb^{(0)} +\eps \Pb^{(1)}+\eps^2 \Pb^{(2)}$ by third order terms,
however these third order terms need not be positive semidefinite.
An alternative transformation is 
$ T_b[\Pb^{(\eps)}] \equiv 
\left[\Pb^{(0)} +\eps \Pb_L^{(1)}+\eps^2 \Pb_L^{(2)}\right] \Pb^{(0) -1} 
$ $\left[\Pb^{(0)} + 
\eps \Pb_L^{(1)}+\eps^2 \Pb_L^{(2)}\right]^T$. 
$T_b[\Pb^{(\eps)}]$ approximates $\Pb(\eps)$ only to second order,
but adds only positive terms. Therefore $T_b[\Pb^{(\eps)}]$ can be used to
provide second order upper bounds.

A number of other stabilizing transformations may be defined. 
$\eps \Pb_L^{(1)}$ can be replaced by $\epshl \Pb^{(1)}$
at the cost of losing the $LD^{-1}L^T$ block factorization.
A more useful transformation is to decompose $\Pb^{(\eps)}$ into its
spectral representation, and then to set any negative eigenvalues
of $\Pb^{(\eps)}$ to zero. This spectral transformation has the advantage
that it uses the smallest possible correction which makes the transformed
matrix positive semidefinite. In Appendix A, we describe a first
order approximation to the spectral decomposition.
Instead
of actually performing the singular value decomposition, we may simply test
$\Pb^{(\eps)}$ for negative eigenvalues using Eq. (A3). Only the
eigenvalues with small $\lambda_k^{(0)}$ need be tested and/or replaced.
Thus the eigendecomposition approach is especially attractive
when only a small number of eigenvalues are questionable.

$T_1[\Pb^{(\eps)}]$ also has a block $LDL^T$ representation: 
$[\Ib_N+\eps\Pb_L^{(1)}
\Pb^{(0)-1}]\Pb^{(0)}[\Ib_N+\eps\Pb_L^{(1)} \Pb^{(0)-1}]^T$. 
This factorization is less numerically efficient than
the $LD^{-1}L^T$ representation.

In our filtering applications, we use the transformation,
$\Pb_+^{(\eps)}=T[\Pb^{(\eps)}]$, to stabilize the data assimilation and 
variance evaluations.
We note that $T[T[\Pb^{(\eps)}]] = T[\Pb^{(\eps)}]$. 
This property is important 
when the covariance matrix, $\Pb^{(\eps)}$, is modified many times
with small updates. We let $T[\cd]$ denote the appropriate stabilizing
transformation.

\noindent
{\bf III. DIAGONALLY DOMINANT DISCRETE KALMAN FILTERS}

We consider the discrete linear state space model:
$$
\vec{x}_{i+1} = \Phi (i+1,i) \vec{x}_i + \Ga_i\vec{w}_{i}
,\eqno (3.1)$$
$$
\vec{y}_{i} = \Hb_i \vec{x}_i + \vec{v}_i 
,\eqno (3.2)$$
where $\vec{x}_i$ is the state vector of dimension $N$,
$\vec{y}_i$ is the measurement vector of dimension $m$, and
$\Phi (j,i)$ is the $N \times N$ nonsingular deterministic part of
the map from time $i$ to time $j$. The system noise, $\vec{w}_k$,
is assumed to be an $r$-dimensional white Gaussian with covariance $\Qb_i$.
The measurement noise is a $m$-dimensional white Gaussian sequence
with nondegenerate covariance $\Rb_i$. The $m \times N$ measurement
evaluation matrix, $\Hb_i$, maps the state vector, $\vec{x}_i$, onto
the deterministic part of the measurements. 
We define the $N\times N$ matrices, 
$\Qb_{\Ga,i} \equiv \Ga_i \Qb_i \Ga_i^T$ and  
$\Jb_{i} \equiv \Hb_i^T \Rb_i^{-1} \Hb_i$.

The standard Kalman filter estimates the state vector, 
$\hat{\vec{x}}(i|j)$, at time $i$ given
the measurements, $\vec{y}_1 , \ldots , \vec{y}_j$ up to time $j$
by the time evolution update:
$$
\hat{\vec{x}}(i+1|i) = \Phi (i+1,i) \hat{\vec{x}}(i|i)\ . \eqno (3.3)
$$
The covariance, $\Pb(i|j)$,  of the estimate, $\hat{\vec{x}}(i|j)$,
evolves as
$$
\Pb(i+1|i) = \Phi (i+1,i) \Pb(i|i) \Phi^T (i+1,i) + 
\Qb_{\Ga,i}  \ . \eqno (3.4)
$$
We assume that $\hat{\vec{x}}(0|0)$ and $\Pb(0|0)$ are given.
The measurement update is
$$
\hat{\vec{x}}(i|i) = \hat{\vec{x}}(i|i-1) + \Kb_i ( \vec{y}_i -
\Hb_i \hat{\vec{x}}(i|i-1) ) 
,\eqno (3.5)$$
$$
\Pb(i|i)^{-1} = \Pb(i|i-1)^{-1} + \Hb_i^T \Rb_i^{-1} \Hb_i 
,\eqno (3.6)$$
where $\Kb_i$ is the $N \times m$ Kalman gain:
$$
\Kb_i = [\Pb(i|i-1)^{-1} + \Hb_i^T \Rb_i^{-1} \Hb_i ]^{-1} \Hb_i^T \Rb_i^{-1}
= \Pb(i|i) \Hb_i^T \Rb_i^{-1}
. \eqno (3.7)$$
We now assume that $\Phi (i+1,i)$, $\Qb_{\Ga,i}$,
$\Pb(0|0)$ and $\Jb_i \equiv \Hb_i^T \Rb_i^{-1} \Hb_i^T$ 
are N.B.D. We assume that the leading order operator, $\Phi(i+1,i)$,
is normal, so that its eigenvectors are orthogonal.
For clarity, we denote
$\Phi^{(0)}(i+1,i)$ by $\Lambda_i$.
We denote the $k, \ell$-th subblocks of $\Pb(i|j)$ by
$\Pb(i|j)_{\{ k, \ell\}}$, and use similar subscripts for the subblocks
of $\Jb_i$, $\Qb_{\Ga,i}$ etc.
We present the expansion of the Kalman filter only to first order. Higher
order expressions are similar, but longer. 

 
The time evolution update for
$\hat{\vec{x}}$, Eq. (3.3), may be computed to arbitrary order if desired.
The time evolution of the covariance  for the standard N.B.D.
representation satisfies
$$
\Pb^{(0)}(i+1|i)_{\{k,k\}} = \Lambda(i)_{k} \Pb^{(0)}(i|i)_{\{k,k\}}
\Lambda(i)_{k}^T + [\Ga\Qb\Ga^T]^{(0)}_{ i \{k,k \}}
,\eqno (3.8)$$
for the zeroth order block diagonal covariance and
$$
\Pb^{(1)}(i+1|i)_{\{k,\ell\}} = 
\Lambda(i)_{k} \Pb^{(1)}(i|i)_{\{k,\ell\}}  \Lambda(i)_{\ell}^T +
\Phi^{(1)}(i+1,i)_{\{k,\ell\}} \Pb^{(0)}(i|i)_{\{\ell,\ell\}}
  \Lambda(i)_{\ell}^T 
$$
$$ +
\Lambda(i)_{k} \Pb^{(0)}(i|i)_{\{k,k\}}\Phi^{(1)}(i+1,i)_{\{\ell,k\}}^T
+ [\Qb_{\Ga,i}]^{(1)}_{\{k,\ell\}}
, \eqno (3.9) $$
for the first order covariance.
The measurement update of $\hat{\vec{x}}(i|i)$ and $\Pb(i|i)$ is
separated into four steps. First, the zeroth order, block diagonal
approximation to $\Pb(i|i)$ is determined by solving the block system
$$
\Pb^{(0)}(i|i)= [\Pb^{(0)}(i|i-1)^{-1} + \Jb^{(0)}_{i} ]^{-1}
\eqno (3.10)
$$
exactly. Second, the first order corrections to $\Pb(i|i)$ are
$$ \Pb^{(1)}(i|i)= 
\Pb^{(0)}(i|i)^{-1} [\Pb^{(0)}(i|i-1)^{-1}
\Pb^{(1)}(i|i-1) \Pb^{(0)}(i|i-1)^{-1} - \Jb^{(1)} ] 
\Pb^{(0)}(i|i)^{-1} 
.\eqno (3.11)$$

Third, our estimate of $\Pb(i|i)$ is forced to be positive semidefinite
using the transformation:
$\Pb^{(\eps)}_{+}(i|i) = T[\Pb^{(\eps)}(i|i)]$.
Finally, we update our estimate of $\vec{x}(i|i)$ using
$$
\hat{\vec{x}}(i|i) = \hat{\vec{x}}(i|i-1) + \Pb_+^{(\eps)}(i|i)
\Hb_i^T \Rb_i^{-1} [ \vec{y}_i - \Hb_i \hat{\vec{x}}(i|i-1) ] 
, \eqno (3.12)$$
where the data assimilation is evaluated exactly.
In our stabilized filter, the stabilizing terms,
$T[\Pb^{(\eps)}(i|i)]-\Pb^{(\eps)}(i|i)$, are not propagated in the
filter. Instead, new stabilizing terms are calculated at every data 
assimilation step. If $\Pb(i+1|i)$ or $\Pb(i|i)$ needs to be evaluated
to assess the uncertainty in the state space estimate, we use the
stabilized approximations.

Comments:


1) Our covariance matrices, $\Pb^{(\eps)}(i|i)$, are approximations
of the covariance matrices of the optimal estimate, $\xvh(i|i)$,
and not the actual covariance of the approximate estimate,
$\xvh^{(\eps)}(i|i)$. 

2) Computational savings occurs only for the 
first order approximation to the state vector covariance matrix, $\Pb$, 
and not for the state estimate. 
The computational requirements of second order calulations are approximately
equal to the costs of the original Kalman filter.
Our approximate filter does not require successive matrix inversions,
and is therefore more numerically stable.

3) The zeroth order block matrices, $\Pb^{(0)}(i|i)$ and $\Pb^{(0)}(i|i-1)$,
are positive definite since the matrix operations are performed exactly
on the each separate block of the zeroth order matrix.

4) Positive definite reformulations of the
Kalman filter such as Potter's algorithm, square root filtering and UD
filtering (Bierman (1977)) are not often unnecessary since the transformations,
$\Pb \rightarrow T[\Pb]$, guarantee positive semidefiniteness.

5) In general, the block diagonal structure is incompatible with
sequential processing of the measurements
since each $\Hb_{i, \{ k \}} \Rb^{-1}_{ \{ k,k \} } \Hb_{i \{ k \} }$
separately is usually not block diagonal.

6) Suboptimal versions of the information filter reformulation
may be constructed using duality. A second order upper bound on
$\Pb^{-1}$, constructed using an information filter and the stabilizing
transformation, $T_b$, and thereby producing 
a lower bound on $\Pb$.

7) Different suboptimal filters with positive definite covariance may be
constructed by expanding other formulations of the Kalman filter order
by order and inserting the transformation $\Pb \rightarrow T[\Pb]$ whenever
necessary.
To guarantee that the N.B.D. structure is fully utilized,
each matrix in the reformulation should be N.B.D. For
example, replacing Eq. (3.6) by 
$\Pb(i|i) = \Pb(i|i-1) -\Kb_i \Hb_i \Pb(i|i-1),$ 
or 
$\Pb(i|i) = [\Ib_N - \Kb_i \Hb_i ]\Pb(i|i-1) [\Ib_N-\Kb_i \Hb_i ]^T 
+ \Kb_i \Rb_i \Kb_i^T$ 
yields a system of equations where each term in the evaluations
is not explicitly N.B.D.
Similarly, replacing the Kalman gain matrix, $\Kb_i$, of Eq. (3.7) with
the representation, 
$\Kb_i \equiv \Pb(i|i-1) \Hb_i^T [\Hb_i \Pb(i|i-1) \Hb_i^T + \Rb_i ]^{-1}$,
which is not in N.B.D. form,
results in a system which is not explicitly N.B.D.

\noindent
{\bf IV. N.B.D. FIXED INTERVAL SMOOTHERS}

In this section, we derive suboptimal, second order approximations
to the various formulations of the fixed interval Kalman smoother.
We denote the final measurement time by $N_f$. 
We begin with the Rauch-Tung-Striebel (R.T.S.) formulation of the smoother.  
We then present a new information formulation of the R.T.S. smoother
as well as the Bryson-Frazier formulation. 
The R.T.S. smoother consists of 
a forward Kalman filter followed by a backward smoother correction. 
This structure arises because the estimation equations
for $\vec{x} (i|N_f)$ have a block tribanded structure. The 
forward-backward sweeps correspond to the standard algorithm 
for solving block
tribanded matrices. In our notation, the R.T.S. smoother 
(Rauch et al. (1965), Bryson and Ho, Ch. 13.2 (1969)) is
$$
\hat{\vec{x}}(i|N_f) = \hat{\vec{x}}(i|i)+
\Pb(i|i) \Phi (i+1|i)^T \Pb^{-1}(i+1|i) 
\left( \xvh(i+1|N_f)- \xvh(i+1|i) \right) 
, \eqno (4.1) $$
$$ \Pb(i|N_f) = \Pb(i|i) + \eqno (4.2) $$
$$
\Pb(i|i) \Phi(i+1|i)^T \Pb^{-1}(i+1|i) 
\left[ \Pb(i+1|N_f)- \Pb(i+1|i) \right]  
\Pb^{-1}(i+1|i) \Phi(i+1|i)\Pb(i|i) 
\ . $$
We assume that $\Pb(i,i)$ and $\Pb(i+1,i)$ have been computed using
the N.B.D. approximations and stabilizing transformations of Sec. II.
We stabilize both $\Pb(i|i)$ and $\Pb^{-1}(i+1|i)$ before evaluating
Eq. (4.1) to all orders. 
The R.T.S. fixed interval smoother is explicitly in
N.B.D. form, and the N.B.D. expansion of Secs. II and III
is used to evaluate $\Pb(i|N_f)$ to second order. 
To ensure positive definiteness, we stabilize
our estimate of $\Pb(i|N_f)$:
$\Pb^{(\eps)}_+(i|N_f) = T[\Pb^{(\eps)} (i|N_f)]$.

A desirable property of a smoother is that 
$\Pb^{(\eps)}(i|i)\ge  \Pb^{(\eps)}(i|N_f)\ge \ 0 $,
and unfortunately our suboptimal approximation of the R.T.S.
smoother does not explicitly insure this property for moderate
values of $\eps$. In contrast,
we now show the information formulation of
the R.T.S. smoother covariance equation possesses the property that
$\Pb^{-1(\eps)}(i|N_f) \ge  \Pb^{-1(\eps)}(i|i)\ge \ 0 $.
We apply the Sherman-Morrison matrix inverse identity to
Eq. (4.2) twice and simplify to yield

$$\Pb^{-1}(i|N_f) = \Pb^{-1}(i|i) + 
 \Phi(i+1|i)^T 
\left[ 
\left( \Pb^{-1}(i+1|N_f)- \Pb^{-1}(i+1|i) \right)^{-1}   
+ \Qb_{\Ga,i} \right]^{-1}
\Phi(i+1|i) 
.\eqno (4.3) $$
We construct a suboptimal smoother by expanding the estimation
equations in powers of $\epsilon$ and applying the stabilizing transformation,
$\Pb^{-1(\eps)}(i|N_f) \rightarrow T[\Pb^{-1(\eps)}(i|N_f)].$ 

The original R.T.S. formulation requires that the
evolution equations be integrated backward in time during the backward
sweep. Since we are interested in distributed dynamical systems with
dissipation and diffusion, such a backward integration is ill-conditioned. 
The Bryson-Frazier formulation of the fixed interval smoother 
reduces this problem by making the following change of variables
for the smoother correction:
$$
\hat{\vec{x}}(i|N_f) = \hat{\vec{x}}(i|i)-\Pb(i|i) \Phi (i+1|i)^T
\lambda (i) 
,\eqno (4.4)$$
$$
\Pb(i|N_f) = \Pb(i|i)-\Pb(i|i) \Phi (i+1|i)
\La{(i)} \Phi (i+1|i)^T \Pb(i|i) 
.\eqno (4.5)$$
In terms of the
auxiliary $N$ vector $\vec{\lambda} (i)$ and $N \times N$ positive
definite symmetric matrix ${\bf \La}(i)$, equations (4.1)-(4.2) transform
to
$$
\vec{\lambda} (i-1) = (\Ib_n - \Pb(i|i)\Jb_i )^T [ \Phi (i+1,i)
\vec{\lambda} (i) - \Hb_i^T \Rb_i^{-1} ( \vec{y}_i - \Hb_i \hat{\vec{x}}
(i|i-1))] 
,\eqno (4.6)$$
$$
{\bf \La}(i-1)=(\Ib_n -\Pb(i|i)\Jb_i )^T
\Phi (i+1,i)^T {\bf \Lambda} (i) \Phi (i+1,i)(\Ib_n - \Pb(i|i)\Jb_i )
+\Jb_i -\Jb_i \Pb(i|i) \Jb_i ,\eqno (4.7)
$$
subject to the final conditions:
$\vec{\lambda} (N_f)= 0 \ ,  \La(N_f) = 0 $.
$\La(i)$ is positive semidefinite, but its approximation, $\La^{(\eps)}(i)$,
need not be.
We do not stabilize our estimate of $\La^{(\eps)}(i)$ 
or any term in Eq. (4.7),
since adding positive definite terms to $\La(i)$ will tend to
underestimate $\Pb(i|N_f)$. Instead, we again apply the stabilizing
transformation only to $\Pb(i|N_f)$: $\Pb(i|N_f)\rightarrow T[\Pb(i|N_f)]$. 
 
\ns \noindent
{\bf V. CONCLUSION}

In this article, we have given first and second 
order  approximations for the Kalman filter and a number of smoothers
by expanding the estimation equations in powers of the coupling parameter. 
We have described the formulations the estimation equations 
which explicit preserve the N.B.D. structure. 
We apply stabilizing transformations to ensure the approximate covariance
is positive semidefinite. We do not propagate these stabilizing terms in the 
Kalman filter in order to minimize the perturbation. 

Other N.B.D. formulations are possible where stabilizing terms are added
to the covariance and propagated in the filter. To minimize the effect of the
terms the approximate spectral representation of Eq. (A5) may be used.
Only the small or negative eigenvalues need be modified.  
The stabilizing transformation need not be applied at every time step. 
Instead, the values of $\Pb(i|i)$ may be examined occasionally or regularly,
and stabilized if they have eigenvalues near zero.

The computational advantage in reducing the operations count by using first 
order approximations is apparent and scales as $O(1/N_b)$. If the 
stochastic system has a special structure like nearest neighbors block
structure, computational savings may also be present 
for second order approximations.  

For general N.B.D. structure, 
second order approximation actually increases the computational
work over straightforward, nonexpansion calculations.
In spite of the additional complication and computational cost,
higher order calculations are sometimes necessary and valuable. 
Higher order calculations for weakly coupled systems have been
given in Shen and Gajic (1990).
To motivate
second order approximations, we consider a case where $\Pb^{(0)}$
is the identity matrix and that $\Pb^{(1)}(\eps)$ has a large
negative eigenvalue, $\la^{(1)}_1$, such that  
$\eps^* \la^{(1)}_1 = -1 +\delta^*$ for the value of $\eps^*$ of interest.
$\Pb^{(0)}+ \eps^*\Pb^{(1)}(\eps^*)$ will have at least one small eigenvalue, 
$\la_1 \sim \delta^*$, and a corresponding large eigenvalue,
$O({1/\delta^*})$, for its inverse.
Our approximate filter effectively replaces this large matrix component 
by terms of order $O(1/4)$. 
Thus we have increased the stability at the cost
of accuracy and slightly longer computational time. The enhancement
in numerical stability will be greatest when $\Pb^{(0)}$ is
bounded from below and $\Pb^{(0)}+ \eps^*\Pb^{(1)}(\eps^*)$ 
is close to singular.   



\noindent
{\bf Acknowledgement}

We thank the referees for their helpful comments.
This work was supported by the U.S. Department of Energy, Grant No.
DE-FG02-86ER53223.
\vspace{.1in}

\begin{center}
{\bf References}
\end{center}

\ni
Anderson, B.D.O. and J.B. Moore (1979). {\it Optimal Filtering.}
Prentice-Hall, New Jersey.

\ni
Bierman, G.J. (1977). {\it Factorization Methods for Discrete
Sequential Estimation}. Academic Press, New York.

\ni
Bryson, Jr., A.E. and Y.C. Ho (1969). {\it Applied Optimal Control}.
Blaisdel Publishing Co., New York.

\ni
Cohn, S.E. and D.F. Parrish (1991). The behavior of
forecast error covariances for a Kalman filter in
two dimensions.
{\it Monthly Weather Review} {\bf 120}, 1757-1785.

\ni
Canuto, C.,  M.Y. Hussaini, A. Quarteroni, and T.A.Zang (1988).
{\it Spectral methods in fluid dynamics.} 
Springer-Verlag, New York.

\ni
Gajic, Z., D. Petkovski, X. Shen (1990). {\it Singularly perturbed
and weakly coupled linear control systems.} Lecture Notes in Control and
Information Sciences, No. 140. Springer-Verlag, Berlin.

\ni
Jazwinski, A.H. (1970). {\it Stochastic Processes and
Filtering Theory}. Academic Press, New York.

\ni
Kokotovic, P., W. Perkins, J. Cruz Jr. and G. D'Ans (1969).
$\epsilon$-coupling for near-optimum design of large scale linear
systems. {\it Proc. IEEE} {\bf 116}, 889-892.

\ni
Moore, J.B. (1973).
Discrete time fixed lag smoothing. {\it Automatica}
{\bf 9}, 163.

\ni
Rauch, H.E., F. Tung and C.T. Striebel (1965). Maximum likelihood
estimates of linear dynamic systems. {\it AIAA J.} {\bf 3}, 1445.

\ni
Sezer, M. and D. Siljak (1986). Nested $\epsilon$-decomposition and
clustering of complex systems. {\it Automatica} {\bf 22}, 321-331.

\ni
Shen, X.-M. and Z. Gajic (1990). Near-optimum steady state
regulators for stochastic linear weakly coupled systems. {\it
Automatica} {\bf 25}, 919-923.

\ns
\noindent
{\bf APPENDIX A: FIRST ORDER N.B.D. MATRIX OPERATIONS} 

A) {\it Storage and Operations Count}

{ We examine the computational savings which occur when the matrix operations
are performed only to first order in $\eps$.
We define the following moments
$N_2 \equiv \left( N_b \sum_{k=1}^{N_b} n_k^2 \right)^{1/2}$
and $N_3 = \left( N_b^2 \sum_{k=1}^{N_b} n_k^3 \right)^{1/3}$. 
Thus if all blocks are the same size, $n_1 = n_2 = n_{b}$,
$N_2 = N_3 = N$. 
Our operation count is for the number of scalar multiplications.
In contrast to the other sections, we store the first order block diagonal
terms, $\eps\Pb^{(1)}_{\{k,k\} }$, in $\Pb^{(0)}(\eps)$, and assume
$\Pb^{(0)}_{\{k,k\}}(\eps)=\Pb^{(0)}_{\{k,k\}} +\eps\Pb^{(1)}_{\{k,k\} }$
is positive definite. This representation slightly reduces the operation
count. 

We consider two subclasses of N.B.D. matrices: general
and nearest neighbor. General block diagonal matrices have no particular 
structure in $\Pb^{(1)}$ and $\Pb^{(2)}$.
We say a matrix $\Pb^{(\eps)}$ has nearest
neighbor structure if and only if $\Pb^{(1)}$ has nonzero elements only
on the diagonal, $\Pb_{k,k}^{(1)}$ and the adjacent bands,
$\Pb_{k,k \pm 1}^{(1)}$. We say a matrix $\Pb^{(\eps)}$ has a strongly
nearest neighbor structure if and only if
$\Pb^{(2)}$ has nonzero elements only on
the diagonal and two adjacent bands, $\Pb_{k,k \pm 1}^{(2)}$ and
$\Pb_{k,k \pm 2}^{(2)}$ as well.

For the N.B.D. matrices with a general structure on
$\Pb^{(1)}$, the 
storage requirement is $N^2$ for an arbitrary matrix and $N(N+1)/2$
for a symmetric matrix. When the matrix is weakly coupled, the storage
is $\sum_{k=1}^{N_b} n_k^2 + 2n_k n_{k+1}$ for arbitrary matrices and
$\sum_{k=1}^{N_b} {n_k (n_k +1) \over 2} + n_k n_{k+1}$ for
symmetric matrices, where $n_{N_b + 1} \equiv 0$. When all the $n_k$
are equal, $n_1 = n_2 = n_b$, the sums storage requirements 
for nearest neighbor matrices are
$(3N_b -2)n_1^2$ for no symmetry, and
$ ({3 \over 2} N_b -1) n_1^2 + {N_b n_1 \over 2}$ for symmetric matrices.

B) {\it $LD^{-1}L^T$ Factorization}

The transformation $T[\Pb^{(\eps)}]\equiv [\Pb^{(0)} +
\eps \Pb_L^{(1)}] \Pb^{(0) -1} [\Pb^{(0)} + \eps \Pb_L^{(1)} ]^T $ need not
be explicitly computed by multiplying $\Pb_L^{(1)} \Pb^{(0)-1}
\Pb_L^{(1)T}$. Instead, the implicit $LD^{-1}L^T$ representation is
usually sufficient.
 The $LD^{-1}L^T$ factorization requires just ${1 \over 2}
\sum_{k=1}^{N_b} n_k^3$ multiplications to compute $\Pb_0^{-1}$.
The $LD^{-1}L^T$ representation of $T[\Pb^{(\eps)}]$ does
require that both $\Pb^{(0)}$ and $\Pb^{(0)-1}$ be stored. Since
both matrices are symmetric, this requires an additional storage
allocation of $\sum_{k-1}^{N_b} {n_k (n_k +1) \over 2}$.


C) {\it First Order Matrix Multiplication} 

The N.B.D. structure is preserved under matrix
multiplication. Let $\Rb = \Pb\Qb$, then $\Rb^{(0)} =
\Pb^{(0)} \Qb^{(0)}$ and $\Rb^{(1)} =
\Pb^{(0)} \Qb^{(1)} + \Pb^{(1)} \Qb^{(0)}$.
Calculating $\Rb^{(0)}$ requires $\sum_{k=1}^{N_b} n_k^3$ operations and
calculating $\Rb^{(1)}$ requires $2 \sum_{k=1}^{N_b} n_k^2
(N-n_k )$ for a total of $2 {NN_2^2 \over N_b} \ - \
{N_3^3 \over N_b^2}$ operations. If both $\Pb$ and $\Qb$ have
nearest neighbor symmetry, calculating $\Rb^{(1)}$ requires only
$2\sum_{k=1}^{N_B} n_k^2 (n_{k+1} + n_{k-1})$ operations where
$n_0 \equiv 0$, and $n_{N_b +1} \equiv 0$. For equal size blocks, the
total operation count is ${5N^3 \over N_b^2} \ - \ {4N^3 \over N_b^3}$.

A second matrix operation which is often performed in filtering is
${\bf S} = {\bf R QR}^T$, where $\Qb$ is symmetric.
For ordinary matrices, this symmetric product requires 
${3 \over 2}N^3+ {1\over 2}N^2$ operations.
 Computing ${\Sb}^{(0)} = \Rb^{(0)}
\Qb^{(0)} \Rb^{(0)T}$ and $\Rb^{(0)} \Qb^{(0)}
\Rb^{(1) T}$ requires $\sum_{k=1}^{N_b} {n_k^3 \over 2} + (N+{1\over 2})n_k^2 =
{(N+1/2)NN_2^2 \over N_b} + {N_3^3 \over 2N_b^2}$. Estimating $\Rb^{(0)}
\Qb^{(1)} \Rb^{(0) T}$ requires ${3\over2}[{NN_2^2 \over N_b} \ - \
{N_3^3 \over N_b^2}]$ multiplications. Thus the symmetric product
requires a total of ${(5N+1)N_2^2 \over 2N_b} \ - \
{N_3^3 \over N_b^2}$ multiplications. For nearest
neighbor matrices, a total of ${1\over 2}\sum_{k=1}^{N_b} n_k^2 (3n_{k} + 1 + 
{5}(n_{k-1} + n_{k+1}) )$ multiplications are required.
 
D) {\it Matrix Inversion and 
$D^{-1}[D-L]D^{-1} [D - L]^T D^{-1}$ Factorization}


To stabilize the order by order approximate inversion, we define
the $Inv[\cdot]$ transformation to be the $T[\cdot]$ 
transformation applied to the approximate inverse:
$Inv \ [\Pb^{(\eps)} ]$ $ \equiv T[\Pb^{(\eps)-1}]$.
When the approximation is first order, $Inv[\cdot]$ reduces to
$ T[\Pb^{(0)-1} - \epsilon
\Pb^{(0)-1} \Pb^{(1)}\Pb^{(0)-1}]
=\Pb^{(0)-1}[\Pb^{(0)} - \eps \Pb_L^{(1)}]\Pb^{(0)-1}
[ \Pb^{(0)} - \eps \Pb_L^{(1)} ]^T \Pb^{(0)-1}$.
We refer to this factorization of the approximate inverse as the
$D^{-1} L' D^{-1} L'^T D^{-1}$ factorization where
$L'\equiv \Pb^{(0)} - \epsilon \Pb_L^{(1)}$.

The approximate inverse, $Inv[\Pb]$, usually does not need to be
computed explicitly. Instead the $D^{-1}L'D^{-1}L'^{T} D^{-1}$
representation of $Inv_[P^{(\eps)}]$ is defined implicitly.
Given the $L^TD^{-1}L$ representation of $T[\Pb^{(\eps)}]$, 
the $D^{-1} L' D^{-1} L'^T D^{-1}$ representation of
$Inv[T[\Pb^{(\eps)} ]]$ requires no additional storage and no
multiplications. 
The inverse of $Inv$ is $Inv$:
$Inv[Inv[T[\Pb^{(\eps)} ]]] =
T[\Pb^{(\eps)}]$, 
and the $Inv[\cdot]$ operation commutes with the
$T[\cdot]$ operation:
${Inv}[T[\Pb^{(\eps)}]] =
T[{Inv}[ \Pb^{(\eps)}]]$.


E) {\it Inverse Matrix Updates}

In Kalman filtering, we successively update the covariance matrix and
then its inverse. We now examine updates of the $D^{-1}(D-L)
D^{-1}(D-L)^TD^{-1}$ representation under matrix addition. We let
the matrices, $\Mb$ and $\Jb$, be block diagonally dominant 
symmetric matrices with $LD^{-1}L^T$
representations. We wish to derive an $LD^{-1} L^T$ representation
of $\Pb$ where $\Pb^{-1} \equiv \Mb^{-1} + \Jb$. The zeroth
order matrix $\Pb^{(0)}$ satisfies $\Pb^{(0)\ -1} = \Mb^{(0)\ -1}
+\Jb^{(0)}$. Since $\Mb^{(0)\ -1} $ is given in the $LD^{-1}L^T$ factorization,
the computation of $\Pb^{(0)\ -1} $ requires only additions and
no multiplications. However $\Pb^{(0)}$ must then be computed and
this requires ${1\over 2}\sum_{k=1}^{N_b} n_k^3 +n_k^2$ operations. We note that
$\Pb^{(0)\ -1} = \Pb^{-1\ (0)}$, but that
$\Pb^{-1\ (1)} \neq \Pb^{(1)\ -1}$. To determine
$\Pb^{(1)}$, we first determine $\Pb^{-1\ (1)}$, 
and then solve for $\Pb^{(1)}$: 
$$
\Pb^{(1)} = \Pb^{(0)} \Mb^{(0)\ -1} \Mb^{(1)}
\Mb^{(0)-1} \Pb^{(0)} - \Pb^{(0)} \Jb^{(1)} \Pb^{(0)} 
,\eqno (A1)$$
or
$$
\Pb^{(1)} = [\Ib-\Pb^{(0)} \Jb^{(0)\ -1}]
\Mb^{(1)} [\Ib-\Jb^{(0)\ -1}\Pb^{(0)}] -
\Pb^{(0)} \Jb^{(1)} \Pb^{(0)} 
.\eqno (A2)$$
Equation (A2) is better conditioned than Eq. (A1) when
$\Jb^{(0)} \ll \Mb^{(0)-1}$.  
Either formulation requires $\left[ \sum_{k=1}^{N_b}
n_k^2 (3N-2n_k ) \right]$ operations.


F) {\it Approximate Eigenvalues and Eigenvectors} 

The eigenvalues and eigenvectors may be estimated from perturbation theory. We
use the following basic result from linear algebra. Let $\Pb$ be a
symmetric matrix form $\Sb_o + \epsilon \Sb_1$, let $\{ \ev^{(0)}_k \}$
and $\{ \lambda^{(0)}_k \}$ be the eigenvectors and eigenvalues of $\Sb_0$,
then the eigenvalues of $\Sb$ are asymptotically
$$\lambda_k \siml \lambda_k^{(0)} 
+ \epsilon \vec{e}_k^{(0)T} \Sb_1 \vec{e}^{(0)}_k,\eqno (A3) $$
and the eigenvectors are asymptotically
$$
\vec{e}_k  \siml \vec{e}_k^{(0)} - \epsilon (\Sb_o - \lambda_k^{(0)}\Ib)^-
(\Sb_1 - (\vec{e}_k^{(0)T} \Sb_1 \vec{e}_k^{(0)}) \Ib) \vec{e}_k^{(0)}
,\eqno (A4)
$$
where ``-" denotes the Moore-Penrose generalized inverse.
Equation (A3) can be used to track the small eigenvalues under successive updates.
The first order spectral decomposition is given by
$$\Pb = \sum_{k=1}^N \lambda_k \vec{e}_k  \vec{e}^T_k  \siml 
\sum_{k=1}^N (\lambda_k^{(0)} +\eps \lambda_k^{(1)})
(\vec{e}_k^{(0)} + \epsilon \vec{e}_k^{(1)} )
(\vec{e}_k^{(0)} + \epsilon \vec{e}_k^{(1)} )^T
.\eqno (A5)$$
Equations (A3-5) generalize the first order decoupling 
transformation used in weakly coupled systems.

\noindent
{\bf APPENDIX B: N.B.D. FIXED LAG DISCRETE SMOOTHERS}

We consider suboptimal approximations of fixed lag Kalman smoothers 
with N.B.D. structure. 
Moore derived the fixed lag Kalman smoother 
(Moore (1973); Ch. 7.3 of Anderson and Moore (1979)) as the
Kalman filter for the augmented state space,
$\xv_A(t)^T \equiv (\xv(t)^T, \xv(t-1)^T \dots \xv(t-n)^T ) $,
and then simplified the resulting augmented filter. 
In presenting the fixed lag smoother, we rewrite the equations
and reorder the matrix indices of Anderson and Moore to
achieve an explicit N.B.D. structure. We define
$\vec{e}_i^{(1)} = 
[\Ib_N - \Jb_i \Pb(i|i) ] \Hb_i^T \Rb_i^{-1} 
( \vec{y}_i - \Hb_i \xvh(i|i-1) )$ 
and
$\vec{e}_i^{(j+1)} = \Phi (i-j+1,i-j) [\Ib_N - \Pb(i-j|i-j) \Jb_{i-j}]
\vec{e}_i^{(j)}\ .$ 
The fixed lag smoother is
$$\hat{\vec{x}} (i|i+n) = \hat{\vec{x}} (i|i) + \Pb(i|i-1)
\sum_{\ell =1}^n \vec{e}_{i+\ell}^{( \ell +1)} \ . \eqno (B1)$$
The covariance of $\hat{\vec{x}} (i|i+n)$ is
$\Pb(i|i+n) = \Pb(i|i) - $
$$\sum_{\ell =1}^n 
\Pb^{(\ell)}(i+\ell|i+\ell -1) \left( \Jb_{i+ \ell} -
\Jb_{i+ \ell} \Pb(i +\ell|i+\ell)\Jb_{i+ \ell} \right)
\Pb^{(\ell)}(i+\ell|i+\ell-1) 
,\eqno (B2)$$
where
$$
\Pb^{( \ell )} (i+ \ell | i + \ell -1) \equiv \Pb(i|i-1)
\prod_{j=0}^{\ell-1} [\Ib_N - \Jb_{i+j} \Pb(i+j|i+j)] \Phi
(i+j+1,i+j)^T 
.\eqno (B3)$$
To achieve an explicit N.B.D. form, we have replaced
$[\Ib_n - \Hb_{i}^T \Kb_{i}^T ]$ with $[\Ib_n - \Jb_{i} \Pb(i|i) ]$, 
and replaced  $\Hb_i [\Hb_i \Pb(i|i-1)\Hb_i^T + \Rb_i ]^{-1}$ with
$ [\Ib_N - \Jb_i \Pb(i|i) ] \Hb_i^T \Rb_i^{-1}$. 
Alternatively, we could replace
$[\Ib_N - \Jb_{i} \Pb(i|i) ]$ with $ \Pb(i|i-1)^{-1} \Pb(i|i)$
and/or replace
$\Hb_i^T \Rb_i^{-1} - \Jb_i \Pb(i|i)\Hb_i^T \Rb_i^{-1}$ 
by $\Pb^{-1}(i|i-1)\Pb(i|i) \Hb_I^T \Rb_i^{-1}$.
The alternative formulations have the advantages that they involve
fewer matrix multiplications.
Our present formulation has the advantage that $\Pb(i|i-1)$
appears only once in the expression, and that all the other terms are
input quantities, usually known to all orders. 
In the limit that $\Jb_i << \Pb(i|i-1)^{-1}$,
our formulation approximates small terms while the alternative formulation
approximates large terms.
For these reasons, we generally prefer our formulation in Eqs. (B1-3).

We stabilize the data assimilation by using $\Pb^{(\eps)}_+(i-j|i-j)
= T[\Pb^{(\eps)}(i-j|i-j)]$ in evaluating $\vec{e}_i^{(j+1)}$ and 
using $\Pb^{(\eps)}_+(i|i-1)$ in Eq. (B1). If $\Pb(i|i+n)$
is of interest, we also stabilize our approximation of it.
\newpage

Our work is motivated by and generalises the results of Cohn and Parrish.
 In Appendix B of Cohn and Parrish (1991), the authors show that
 if $\Hb_i^T \Rb_i^{-1} \Hb_i$ is diagonal and the evolution equations
 are diagonal, then the estimation covariance will be diagonal.
 In this article, we extend their results to include block diagonal
 systems and Kalman smoothers. More importantly, we relax the requirement of
 exact diagonality and consider small offdiagonal terms. We expand the
 estimation equations in  powers of the offdiagonal terms and develop
 numerically wellconditioned algorithms to compute these approximate estimation

In applying Kalman filtering to global circulation models, Cohn and Parrish
 (Cohn and Parrish (1991))
noted that the evolution equtions are simplest in an eigenfunction basis
while the data assimilation is simplest in a finite difference or finite
element representation. By requiring that the measurement locations
be distributed such that $\Hb_i^T \Rb_i^{-1} \Hb_i$ is nearly block
diagonal, we are able to simplify the data assimilation equation in the
eigenfunction domain.

\end{document}